\documentclass[useAMS,usenatbib]{mn2e}

\usepackage{graphicx}

\newcommand{\mincir}{\raise -2.truept\hbox{\rlap{\hbox{$\sim$}}\raise5.truept
\hbox{$<$}\ }}
\newcommand{\magcir}{\raise -2.truept\hbox{\rlap{\hbox{$\sim$}}\raise5.truept
\hbox{$>$}\ }}
\newcommand{\siml}{\raise -2.truept\hbox{\rlap{\hbox{$\sim$}}\raise5.truept
\hbox{$<$}\ }}
\newcommand{\simg}{\raise -2.truept\hbox{\rlap{\hbox{$\sim$}}\raise5.truept
\hbox{$>$}\ }}
\newcommand{\be}{\begin{equation}}
\newcommand{\ee}{\end{equation}}
\newcommand{\ba}{\begin{eqnarray}}
\newcommand{\ea}{\end{eqnarray}}
\newcommand {\kpc} {$h_{70}^{-1}$ kpc $\;$}

\newcommand {\h} {$h_{70}^{-1}$ Mpc$\;$}
\newcommand {\hh} {$h_{70}^{-1}$ Mpc}
\newcommand {\hhh} {\;h_{70}^{-1} \mathrm{Mpc}}
\newcommand {\ks} {km~s$^{-1} \;$}
\newcommand {\kss} {km~s$^{-1}$}

\newcommand {\mqua} {$\times 10^{14}\;h_{70}^{-1}\;M_{\odot} \;$}
\newcommand {\mquaa} {$\times 10^{14}\;h_{70}^{-1}\;M_{\odot}$}
\newcommand {\mqui} {$\times 10^{15}\;h_{70}^{-1}\;M_{\odot} \;$}
\newcommand {\mquii} {$\times 10^{15}\;h_{70}^{-1}\;M_{\odot}$ }

\newcommand{\degree}{\ensuremath{\mathrm{^\circ}}}
\newcommand{\arcm}{\ensuremath{\mathrm{^\prime}\;}}
\newcommand{\arcs}{\ensuremath{\arcmm\hskip -0.1em\arcmm \;}}
\newcommand{\arcmm}{\ensuremath{\mathrm{^\prime}}}

\title[The puzzling merging cluster Abell 1914]{The puzzling merging cluster 
Abell 1914: new insights from the kinematics of member galaxies}
\author[R. Barrena, M. Girardi and W. Boschin]{R. Barrena$^{1,2}$\thanks{E-mail:
rbarrena@iac.es},
M. Girardi $^{3,4}$ and W. Boschin$^{5}$\\
$^{1}$Instituto de Astrof\'{\i}sica de Canarias, C/V\'{\i}a L\'actea s/n, E-38205 La Laguna (Tenerife), Spain\\
$^{2}$Departamento de Astrof\'{\i}sica, Univ. de La Laguna, Av. del
     Astrof\'{\i}sico Francisco S\'anchez s/n, E-38205 La Laguna
     (Tenerife), Spain\\
$^{3}$Dipartimento di Fisica dell'Universit\`a degli Studi
     di Trieste - Sezione di Astronomia, via Tiepolo 11, I-34143
     Trieste, Italy\\ 
$^{4}$INAF - Osservatorio Astronomico di Trieste,
     via Tiepolo 11, I-34143 Trieste, Italy\\ 
$^{5}$Fundaci\'on G. Galilei - INAF (Telescopio Nazionale Galileo), 
     Rambla J. A. Fern\'andez P\'erez 7, E-38712 Bre\~na Baja (La Palma), Spain}
\begin{document}

\date{Accepted DATE. Received DATE; in original form DATE}

\pagerange{\pageref{firstpage}--\pageref{lastpage}} \pubyear{2012}

\maketitle

\label{firstpage}

\begin{abstract}
We analyze the dynamical state of Abell 1914, a merging cluster
hosting a radio halo, quite unusual for its structure. Our study
considers spectroscopic data for 119 galaxies obtained with the
Italian Telescopio Nazionale Galileo. We select 89 cluster members
from spatial and velocity distributions. We also use photometry
Canada-France-Hawaii Telescope archives. We compute the mean cluster
redshift, $\left<z\right>=0.168$, and the velocity dispersion
  which shows a high value, $\sigma_{\rm V}=1210_{-110}^{+125}$ \kss.
From the 2D analysis we find that Abell 1914 has a NE-SW elongated
structure with two galaxy clumps, that mostly merge in the plane of
the sky.  Our best, but very uncertain estimate of the velocity
dispersion of the main system is $\sigma_{\rm V,main}\sim 1000$
\kss. We estimate a virial mass $M_{\rm sys}=1.4$--2.6 \mquii for the
whole system. We study the merger through a simple two-body model and
find that data are consistent with a bound, outgoing substructure
observed just after the core crossing. By studying the 2D distribution
of the red galaxies, photometrically selected, we show that Abell 1914
is contained in a rich large scale structure, with two close companion
galaxy systems, known to be at $z\sim 0.17$. The system at SW supports
the idea that the cluster is accreting groups from a filament aligned
in the NE-SW direction, while that at NW suggests a second direction
of the accretion (NW-SE). We conclude that Abell 1914 well fits among
typical clusters with radio halos.  We argue that the unusual radio
emission is connected to the complex cluster accretion and suggest
that Abell 1914 resembles the well-known nearby merging cluster Abell
754 for its particular observed phenomenology.

\end{abstract}


\section{Introduction}

A fraction of galaxy clusters shows the presence of diffuse
radio emission on Mpc scale. In general, we can distinguish between
two morphologies: ``radio halos'' and ``radio relics''. In the first
case, the emission comes from central cluster regions, while
``relics'' take place in the peripheral zones
\citep{gio02,fer08,ven11,fer12}. The synchrotron origin of this radio
emission reveals the presence of a large-scale magnetic field
and relativistic particles spread out of the cluster. Nowadays,
cluster mergers seem to be the most reasonable framework proposed to
provide enough energy for accelerating electrons to relativistic
velocities and for magnetic field amplification. In this scenario,
radio relics seem to be directly linked with merger shocks
\citep{ens98,roe99,ens01,hoe04}. Instead, the turbulence
following cluster mergers has been proposed as one of the most
important effects to produce giant radio halos
\citep{bru01,bru09}. However, the precise scenario for radio halo
formation is still debated. In fact, there are two main
theoretical approaches to the problem: re-acceleration vs. hadronic
models (Brunetti et al. \citeyear{bru09} and refs. therein). 


Usually, X-ray observations are used to study the dynamical state of
clusters with diffuse radio emission. Indeed, all statistical
analyses are derived from X-ray data \citep{sch01,buo02,cas10,ros11}
and properties of radio emission are derived in general from X-ray
temperature and luminosity \citep[see e.g.][and references
  therein]{gio02}.  In fact, predictions based on turbulent
re-acceleration models well agree with the radio observations of
halos \citep{cas06}. In this sense, \cite{gov01} also find a strong
correlation between X-ray and radio emission when they compare
point-to-point individual surface brightnesses. In addition,
\citet{bas12} reveal no evidence of bimodality in the radio-power --
integrated SZ effect diagram while, at the contrary, \citet{bru07}
find this bimodal feature in the radio-power -- X-ray luminosity
diagram. This fact reveals the need to manage other research ways in
addition to X-ray techniques.

Optical information is an important way to investigate the dynamics of
cluster mergers \citep{gir02}. The spatial distribution and kinematics
of galaxy members allow us to detect substructures and to analyze
possible pre- and post-merging groups, and to distinguish
between evolving mergers and remnants. Moreover, optical data are
complementary to X-ray information because the ICM and galaxies
react on different timescales during a collision. This is clearly
shown in numerical simulations by \cite{roe97}. Thus, for example, the
importance of combining X-ray and optical data to study merging
scenarios is shown by MUSIC (MUlti-Wavelength Sample of Interacting
Clusters) project \citep{maur11}.

In this context, we are now progressing on the DARC (Dynamical
Analysis of Radio Clusters, see \citealp{gir10}\footnote{see also
  http://adlibitum.oat.ts.astro.it/girardi/darc, the web site of the
  DARC project.}) project, which uses spectroscopic and photometric
information of galaxy members to analyze the internal dynamics of
clusters with diffuse radio emission.

We have carried out an intensive observational program focused on the
cluster of galaxies Abell 1914 (hereafter A1914). A1914 is a rich
cluster, X-ray luminous, hosting a hot ICM. It shows an Abell richness
class $R=2$ \citep{abe89}, $L_\mathrm{X}$(0.1--2.4 keV)=17.93$\times
10^{44}\ h_{50}^{-2}$ erg\ s$^{-1}$ \citep{ebe96} and $kT_{\rm X}\sim
9$ keV \citep{bal07,mau08}. Following the Bautz-Morgan classification,
A1914 is a type II structure \citep{abe89}, while it is a
``L-type'' (``linear'') cluster in the Rood-Sastry morphological
scheme \citep{str87}.

\cite{dah02} study the mass and light distributions using weak-lensing
techniques. They recover the elongated shape of this cluster in the
NE-SW direction and find that the light distribution well follows the
mass profile. The two brightest cluster galaxies trace the two
highest peaks in the mass distribution, although in the reverse order,
that is, the highest peak is close to the second brightest galaxy 
\citep{oka08}. On the other hand, \cite{jon05} analyze the galaxy
distribution in the POSS digital. They find signs of dynamical activity 
with two distinct groups of galaxies with no single dominant galaxy.

\cite{buo96} develop the first analysis of the X-ray morphology 
of this cluster using ROSAT data. They find that A1914 is a relaxed
structure, but \cite{jon05} show some evidence against this thesis,
suggesting that this cluster is not so relaxed. They find no 
evidence for a cool core, an unusual high X-ray temperature,
and notice that ROSAT data are very poorly fitted using a
$\beta$ model. Then, using Chandra X-ray data, \cite{gov04} show clear
evidence of merger. In fact, they find a clear elongation of the 
X-ray surface brightness (along WNW-ESE, see Fig.~5d of Govoni et
al. \citeyear{gov04}). In the last years, using Chandra data, A1914 has
been classified as a non relaxed cluster \citep{bal07,mau08}.

Concerning the radio emission, \cite{kom94} first report evidence for
a diffuse and extended radio source \citep[see also][]{gio99,kem01}.
Moreover, \cite{bac03}, using VLA data, show the presence of a 
unpolarized halo. The halo covers a $7.4^{\prime} \times
5.3^{\prime}$ area with a power $P_{\rm 1.4\,GHz}=
8.72\times10^{24}\ h_{70}^{-2}$ W\ Hz$^{-1}$. \citet{gov04}
point out that the diffuse radio emission is quite puzzling with a
bright component elongated in the NW-SE direction and a more typical
low-brightness halo in the cluster center (see Fig.~5d of Govoni et
al. \citeyear{gov04}). The bright radio region does not follows 
either the elongation of the X-ray surface brightness. This fact 
is quite unusual, because in the majority of clusters the elongated 
diffuse radio halo follows the direction of the merger (e.g., the 
``Bullet'' Cluster 1E0657-56, Markevitch et al. \citeyear{mar02}; 
but see Abell 523, Giovannini et al. \citeyear{gio11}).

Despite several studies based on X-ray data, published redshift data
are not enough to perform the detailed dynamical study of A1914.
The work we present here is based on new spectroscopic data
obtained with the Telescopio Nazionale Galileo (TNG). We also use 
photometric data from the Canada-France-Hawaii Telescope (CFHT)
archive.

This paper is organized as follows. We present optical data, including
redshifts and photometry information, in Sect.~2. We expose our
results on the cluster structure in Sect.~ 3. The discussion on the 
dynamical state of A1914 and conclusions are presented in 
Sect.~4 and ~5, respectively.  

\begin{figure*}
\centering 
\includegraphics[width=18cm]{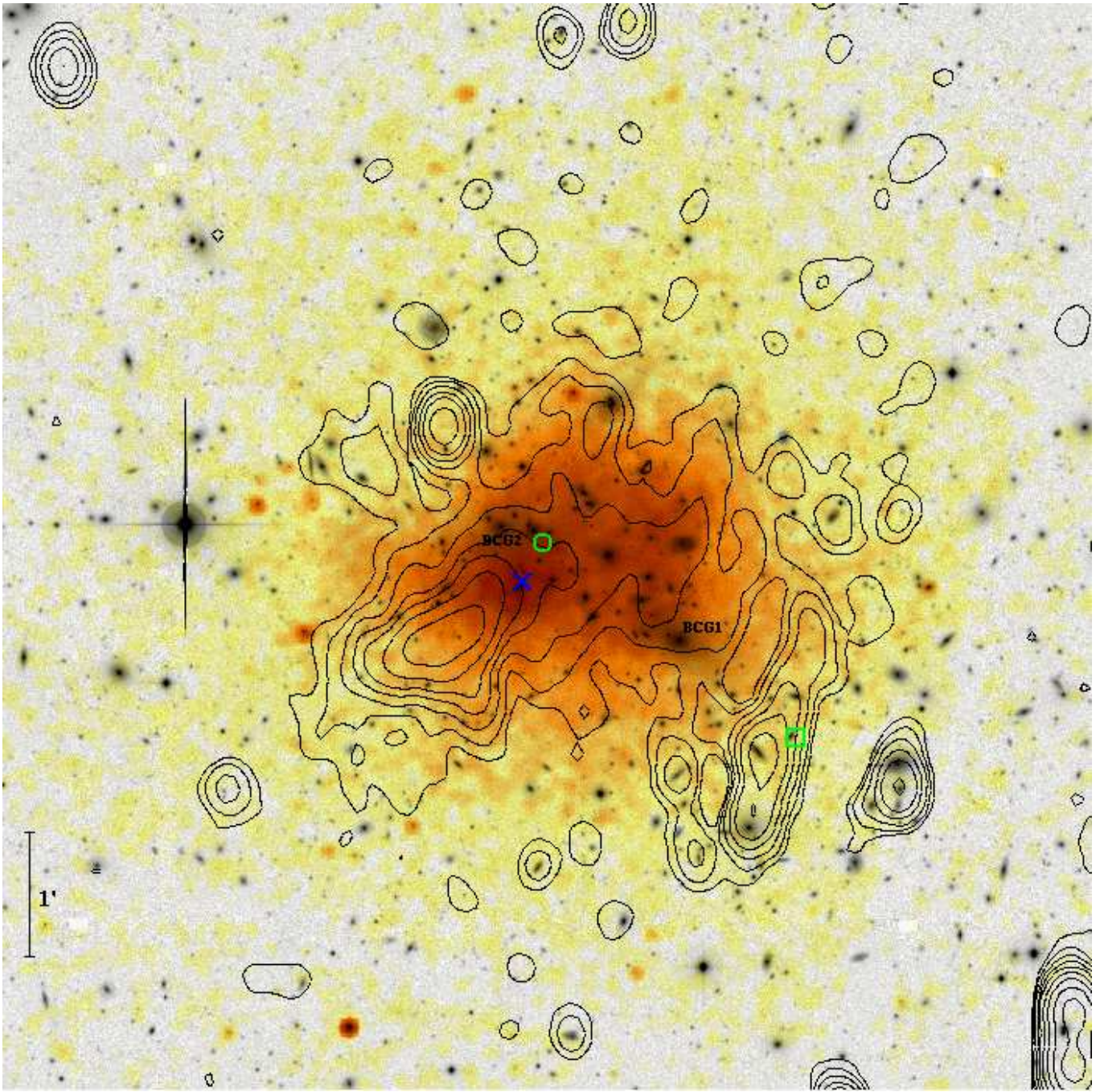}
\caption{Multiwavelength image of the central region of
A1914. The gray-scale image in background corresponds to the
optical $r^{\rm Mega}$-band (CFHT archive). Superimposed, with
orange and yellow colors, we also show the smoothed X-ray image in
the 0.3-7 keV energy range (Chandra archive). Contour levels
represent the VLA radio image at 1.4 GHz (courtesy of F. Govoni; see
Bacchi et al. \citeyear{bac03}). Green circle and square mark the 
centers of the density peaks detected in our analysis of the galaxy
distribution (see Sect.~\ref{photo}). Blue "X" marks the centroid of 
the X-ray surface brightness (Govoni et al. 2004). Labels
indicate the positions of the two brightest cluster galaxies 
(BCG1 and BCG2; see Sect.~\ref{cat} and Table~\ref{catalogA1914}). 
North is up and East is left.}
\label{figimage}
\end{figure*}
  
Unless otherwise stated, we present errors at the 68\% confidence
level (hereafter c.l.). Along this paper, we work using $H_0=70$ km
s$^{-1}$ Mpc$^{-1}$ and $h_{70}=H_0/(70$ km s$^{-1}$ Mpc$^{-1}$) and a
flat cosmology with $\Omega_0=0.3$ and $\Omega_{\Lambda}=0.7$. Within
this cosmology, 1\arcm corresponds to $\sim 172$ \kpc at the cluster 
redshift.

\begin{figure*}
\centering 
\includegraphics[width=18cm]{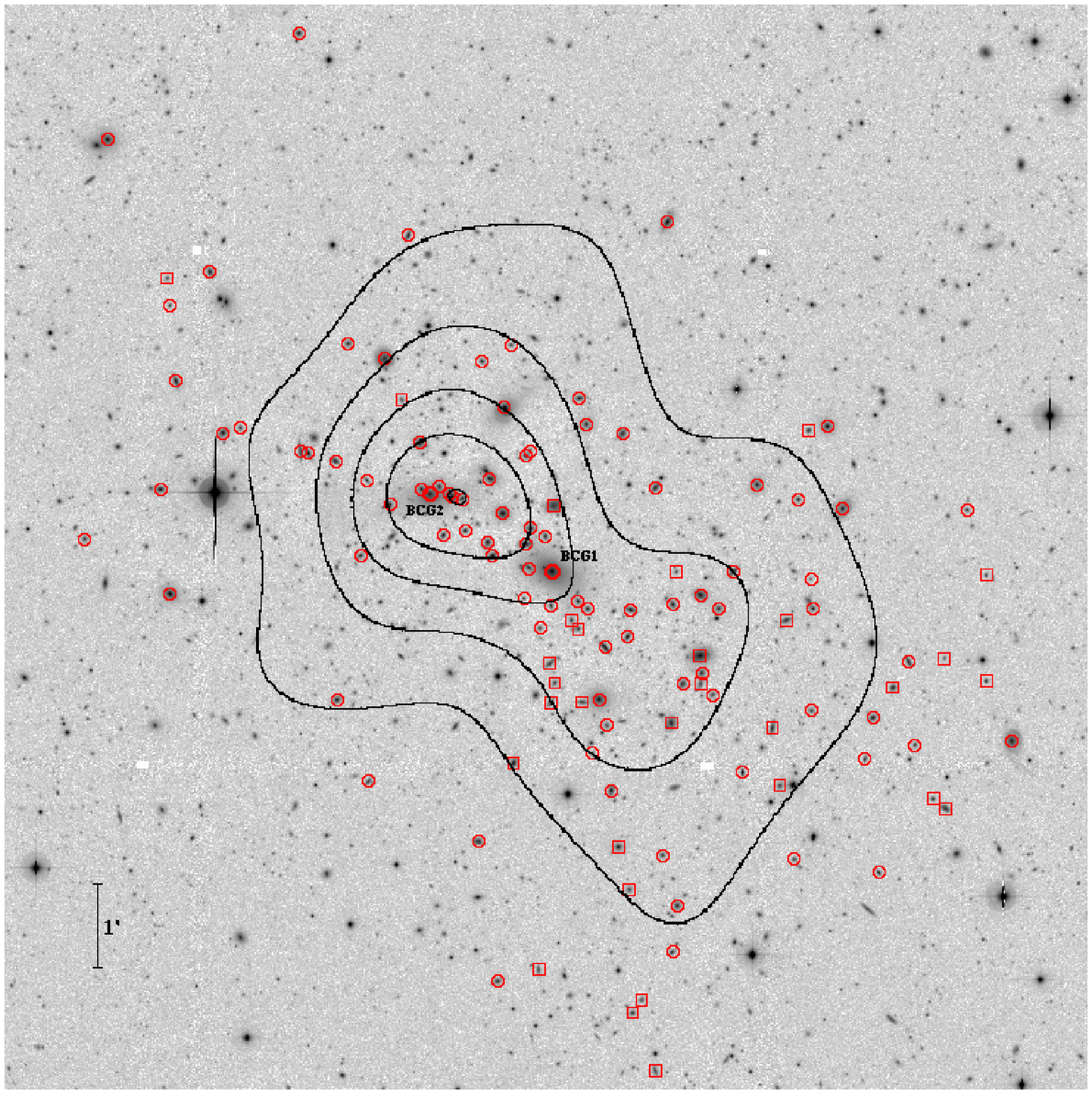}
\caption{$r^{\rm Mega}$-band image of Abell 1914 (CFHT archive).
Circles and squares correspond to galaxy members and
nonmembers, respectively (see Table~\ref{catalogA1914}). Labels
and enhanced circles mark the positions of the two brightest cluster 
galaxies (BCG1 and BCG2; see Sect.~\ref{cat} and Table~\ref{catalogA1914}). 
Black contours are the isodensity contours of the distribution of 
likely member galaxies (see Sect.~3.6 and Fig.~9-top panel). North 
is up and East is left.}
\label{figottico}
\end{figure*}

\section{The data sample}
\label{data}

\subsection{Spectroscopic data}
\label{spec}

We performed observations of A1914 using DOLORES multi-object
spectrograph at the TNG telescope in March 2010. We used the LR-B
grism, which provides a dispersion of 187 \AA/mm. DOLORES works
with a $2048\times2048$ pixels E2V CCD. The pixel size is 13.5
$\mu$m. We retrieved a total of 4 MOS masks containing 146
slits. We exposed 3600 s for each mask.

Spectra were reduced using standard IRAF\footnote{IRAF is distributed
by the National Optical Astronomy Observatories, which are operated
by the Association of Universities for Research in Astronomy, Inc.,
under cooperative agreement with the National Science Foundation.}
tasks.  Radial velocities were computed by using the
cross-correlation technique \citep{ton79} with the IRAF/XCSAO
task, as we have proceeded with other clusters already analyzed in
the DARC project (for a detailed description, see e.g. Boschin
et al. \citeyear{bos12}). In six cases (IDs.~78, ~79, ~83, ~99, ~100,
~107 and ~112; see Table~\ref{catalogA1914}), we considered the
IRAF/EMSAO redshift (based on the wavelength of emission lines in the
spectra) to get a realistic estimation. So, our catalog lists 113
galaxy redshifts in the field of A1914. We also considered
six redshifts more from the SDSS archive (IDs. from ~114 to ~119; see
Table~\ref{catalogA1914}).

The true intrinsic errors are larger than those formal errors given by
the cross-correlation (e.g., Malumuth et al. \citeyear{mal92};
Ellingson \& Yee \citeyear{ell94}; Quintana et al. \citeyear{qui00};
Bardelli et al. \citeyear{bar94}). To correct this effect, some
galaxies were observed in more than one mask. This allows us to
estimate the intrinsic errors in data of the same quality acquired
with the same instrumentation. Our spectroscopic survey provides
duplicate estimations for 18 galaxies. So, following the method
detailed in \cite{bar09} for these 18 galaxies and using the weighted
mean of the two measurement, we concluded that true intrinsic errors
are larger than formal cross-correlation ones by a factor of two. For
the redshifts estimated with the EMSAO task we considered the
largest value between 100 \kss and the formal error.

We also considered nine galaxies having redshift in the SDSS and
lying in the same field spanned by our spectroscopic data. Three of
these SDSS targets were also observed with the TNG/DOLORES. We
find no systematic deviations between the SDSS and our redshifts. We
add the remaining six galaxies to our TNG catalog. Finally, we
obtain a spectroscopic catalog of 119 galaxies, with a median value of
the $cz$ errors of 74 \kss.

\subsection{Photometry and galaxy catalog}
\label{cat}

We also use public photometric data obtained with
Megaprime/Megacam at the CFHT. In particular, we consider
$g^{\rm Mega}$ and $r^{\rm Mega}$ band\footnote{see the URL
  $http://www2.cadc-ccda.hia-iha.nrc-cnrc.gc.ca/megapipe/docs/proc.html\#photcal$
  for a comparison between Megacam and SDSS filters.} images retrieved
from the CADC Megapipe archive \citep{gwy09}. These images cover an
area of $1.05\times1.16$ deg$^2$ with a deepness of $g^{\rm
  Mega}=27.2$ and $r^{\rm Mega}=26.8$ limiting magnitudes (at the
5$\sigma$ detection level). Megaprime photometry is 90\% 
complete down to $g^{\rm Mega}=24.8$ and $r^{\rm Mega}=24.6$. We corrected 
$g^{\rm Mega}$ and $r^{\rm Mega}$ CFHT magnitudes for galactic 
extinction assuming extinction values obtained from SDSS (DR7) 
in the central cluster region.

Table~\ref{catalogA1914} lists our velocity catalog and photometry 
(see also Fig.~\ref{figottico}). We present an identification (ID) 
number in Col.~1 (galaxy members are listed in italic format); 
right ascension and declination (J2000) in Col.~2; CFHT $r^{\rm Mega}$ 
magnitudes in Col.~3; and heliocentric radial velocities $v=cz_{\sun}$ 
and errors $\Delta v$ in Col.~4 and Col.~5, respectively.

Our spectroscopic sample is 80\% (50\%) complete down to 
$r'$=18.3 (=19.7), within an elongated region of 60 $arcmin^2$ (corresponding
to an area of $1.1\times1.6$ Mpc at the redshift of the cluster) around the 
cluster center.


\begin{table}[!ht]
        \caption[]{Radial Velocities of 119 galaxies in the field of 
	Abel 1914. $\dagger$ and $\ddagger$ denote the BCG1 and BCG2,
          respectively.}
         \label{catalogA1914}
              $$ 
           \begin{array}{l c r r r}
            \hline
            \noalign{\smallskip}
            \hline
            \noalign{\smallskip}

\mathrm{ID} & \alpha,\delta\,(\mathrm{J}2000) & r^{\mathrm{Mega}}\,\,\, & v\,\,\,\,\,& \Delta v\\
 &                  & &\mathrm{(\,km}&\mathrm{s^{-1}\,)}\\
            \hline
            \noalign{\smallskip}  

{\it 01}  & 14\ 26\ 11.65 ,+37\ 55\ 16.6 & 18.44 & 50860 &  50 \\
     02   & 14\ 26\ 19.43 ,+37\ 52\ 24.7 & 20.43 & 55260 & 134 \\
{\it 03}  & 14\ 26\ 19.27 ,+37\ 52\ 05.7 & 20.20 & 50340 & 102 \\
{\it 04}  & 14\ 26\ 16.90 ,+37\ 52\ 29.1 & 18.53 & 50375 &  32 \\
{\it 05}  & 14\ 26\ 18.91 ,+37\ 51\ 12.6 & 18.02 & 52280 &  30 \\
{\it 06}  & 14\ 26\ 24.30 ,+37\ 49\ 21.9 & 19.58 & 49505 &  64 \\
{\it 07}  & 14\ 26\ 19.83 ,+37\ 49\ 56.7 & 17.97 & 50539 &  40 \\
{\it 08}  & 14\ 26\ 16.13 ,+37\ 50\ 36.3 & 18.28 & 49964 &  44 \\
{\it 09}  & 14\ 26\ 15.09 ,+37\ 50\ 39.9 & 19.76 & 49591 &  82 \\
{\it 10}  & 14\ 26\ 05.16 ,+37\ 52\ 54.9 & 18.78 & 52674 &  42 \\
{\it 11}  & 14\ 26\ 08.71 ,+37\ 51\ 39.0 & 19.07 & 51671 &  54 \\
{\it 12}  & 14\ 26\ 11.55 ,+37\ 50\ 23.3 & 18.87 & 48309 & 138 \\
{\it 13}  & 14\ 26\ 11.10 ,+37\ 50\ 22.2 & 18.18 & 46889 &  42 \\
{\it 14}  & 14\ 26\ 09.48 ,+37\ 50\ 16.6 & 18.61 & 50578 &  64 \\
     15   & 14\ 26\ 05.55 ,+37\ 50\ 59.4 & 20.24 & 99106 & 156 \\
{\it 16}  & 14\ 26\ 07.62 ,+37\ 50\ 03.1 & 19.64 & 52026 &  76 \\
{\it 17}  & 14\ 26\ 04.48 ,+37\ 50\ 29.4 & 17.91 & 50718 &  38 \\
{\it 18}  & 14\ 26\ 06.25 ,+37\ 49\ 46.0 & 18.55 & 48881 &  36 \\
{\it 19}  & 14\ 26\ 04.38 ,+37\ 49\ 57.1 & 18.91 & 46491 &  94 \\
{\it 20}\ddagger  & 14\ 26\ 03.89 ,+37\ 49\ 53.4 & 16.39 & 50251 &  54 \\
{\it 21}  & 14\ 26\ 02.75 ,+37\ 49\ 53.9 & 18.51 & 50453 &  48 \\
{\it 22}  & 14\ 26\ 02.57 ,+37\ 49\ 51.3 & 17.44 & 50634 &  40 \\
{\it 23}  & 14\ 26\ 03.08 ,+37\ 49\ 24.7 & 18.58 & 49457 &  38 \\
{\it 24}  & 14\ 25\ 58.19 ,+37\ 50\ 20.2 & 18.72 & 50348 &  36 \\
{\it 25}  & 14\ 26\ 00.46 ,+37\ 49\ 19.7 & 18.28 & 49297 &  42 \\
{\it 26}  & 14\ 26\ 00.17 ,+37\ 49\ 10.2 & 18.14 & 51825 &  40 \\
{\it 27}  & 14\ 25\ 58.19 ,+37\ 49\ 18.5 & 17.90 & 48730 &  50 \\
{\it 28}  & 14\ 25\ 58.05 ,+37\ 49\ 01.4 &  --   & 50168 &  70 \\
{\it 29}\dagger  & 14\ 25\ 56.67 ,+37\ 48\ 59.2 & 15.87 & 51029 &  26 \\
{\it 30}  & 14\ 25\ 57.37 ,+37\ 48\ 19.4 & 19.75 & 51092 &  74 \\
{\it 31}  & 14\ 25\ 55.12 ,+37\ 48\ 38.4 & 19.08 & 48103 &  50 \\
{\it 32}  & 14\ 25\ 54.55 ,+37\ 48\ 33.0 & 19.14 & 49759 &  50 \\
     33   & 14\ 25\ 56.50 ,+37\ 47\ 40.9 & 19.64 & 63968 &  68 \\
{\it 34}  & 14\ 25\ 51.99 ,+37\ 48\ 32.3 & 18.35 & 49702 &  30 \\
{\it 35}  & 14\ 25\ 53.88 ,+37\ 47\ 29.5 & 17.23 & 48844 &  26 \\
{\it 36}  & 14\ 26\ 07.98 ,+37\ 49\ 10.3 & 19.09 & 50582 &  64 \\
{\it 37}  & 14\ 25\ 59.57 ,+37\ 49\ 40.0 & 17.37 & 50669 &  60 \\
{\it 38}  & 14\ 25\ 54.63 ,+37\ 50\ 42.5 & 18.62 & 49378 &  82 \\
{\it 39}  & 14\ 25\ 57.93 ,+37\ 49\ 29.4 & 18.18 & 50600 &  36 \\
            \noalign{\smallskip}			 
            \hline					    
            \noalign{\smallskip}			    
            \hline					    
         \end{array}					 
     $$ 						 
         \end{table}					 
\addtocounter{table}{-1}				 
\begin{table}[!ht]					 
          \caption[ ]{Continued.}
     $$ 
           \begin{array}{l c r r r}
            \hline
            \noalign{\smallskip}
            \hline
            \noalign{\smallskip}

\mathrm{ID} & \alpha,\delta\,(\mathrm{J}2000) & r^{\mathrm{Mega}}\,\,\, & v\,\,\,\,\,& \Delta v\\
 &                  & &\mathrm{(\,km}&\mathrm{s^{-1}\,)}\\

            \hline
            \noalign{\smallskip}

{\it 40}  & 14\ 25\ 57.04 ,+37\ 49\ 24.0 & 19.15 & 52796 &  88 \\
{\it 41}  & 14\ 25\ 56.71 ,+37\ 48\ 35.2 & 18.93 & 50818 &  66 \\
     42   & 14\ 25\ 58.94 ,+37\ 46\ 44.8 & 18.08 & 55204 &  76 \\
     43   & 14\ 25\ 54.86 ,+37\ 47\ 27.5 & 19.04 & 63647 & 134 \\
{\it 44}  & 14\ 25\ 47.88 ,+37\ 48\ 42.7 & 17.70 & 52746 &  34 \\
{\it 45}  & 14\ 25\ 45.91 ,+37\ 48\ 59.0 & 18.25 & 49070 &  44 \\
{\it 46}  & 14\ 25\ 53.16 ,+37\ 46\ 25.4 & 18.40 & 51371 &  30 \\
     47   & 14\ 25\ 49.61 ,+37\ 47\ 12.9 & 17.93 & 55809 &  54 \\
{\it 48}  & 14\ 25\ 39.45 ,+37\ 49\ 43.2 & 17.69 & 50257 &  32 \\
     49   & 14\ 25\ 42.80 ,+37\ 48\ 24.8 & 18.66 & 63410 &  82 \\
{\it 50}  & 14\ 25\ 41.21 ,+37\ 48\ 33.1 & 19.25 & 49330 &  66 \\
{\it 51}  & 14\ 25\ 50.13 ,+37\ 45\ 40.2 & 19.37 & 48884 &  42 \\
     52   & 14\ 25\ 43.62 ,+37\ 47\ 09.4 & 18.52 & 55791 &  78 \\
{\it 53}  & 14\ 25\ 41.30 ,+37\ 47\ 21.5 & 19.60 & 49661 & 154 \\
     54   & 14\ 25\ 43.18 ,+37\ 46\ 28.8 & 19.28 &107888 & 102 \\
{\it 55}  & 14\ 25\ 37.68 ,+37\ 47\ 16.4 & 18.53 & 49832 &  70 \\
{\it 56}  & 14\ 25\ 38.15 ,+37\ 46\ 47.5 & 19.53 & 51154 &  76 \\
{\it 57}  & 14\ 25\ 35.19 ,+37\ 46\ 57.4 & 19.62 & 52210 & 140 \\
{\it 58}  & 14\ 25\ 37.35 ,+37\ 45\ 28.0 & 19.17 & 49115 &  50 \\
{\it 59}  & 14\ 26\ 00.79 ,+37\ 51\ 26.5 & 19.42 & 49776 & 142 \\
{\it 60}  & 14\ 25\ 59.08 ,+37\ 51\ 38.0 & 20.07 & 52324 & 198 \\
{\it 61}  & 14\ 26\ 03.33 ,+37\ 49\ 58.7 & 18.69 & 50782 &  86 \\
{\it 62}  & 14\ 26\ 01.96 ,+37\ 49\ 49.2 &  --   & 48245 &  98 \\
{\it 63}  & 14\ 26\ 01.79 ,+37\ 49\ 27.9 & 19.15 & 49258 &  74 \\
{\it 64}  & 14\ 25\ 55.03 ,+37\ 51\ 00.7 & 18.45 & 49768 &  34 \\
{\it 65}  & 14\ 25\ 52.49 ,+37\ 50\ 35.7 & 18.53 & 48045 &  98 \\
{\it 66}  & 14\ 25\ 58.32 ,+37\ 48\ 40.3 & 19.94 & 50783 & 122 \\
     67   & 14\ 25\ 55.51 ,+37\ 48\ 24.9 & 19.36 & 63942 & 116 \\
     68   & 14\ 25\ 55.10 ,+37\ 48\ 18.9 & 19.65 &136183 & 142 \\
{\it 69}  & 14\ 25\ 52.18 ,+37\ 48\ 13.4 & 18.68 & 50039 &  70 \\
{\it 70}  & 14\ 25\ 49.51 ,+37\ 48\ 36.0 & 18.59 & 50373 &  92 \\
{\it 71}  & 14\ 25\ 53.42 ,+37\ 47\ 11.2 & 19.64 & 49720 & 110 \\
{\it 72}  & 14\ 25\ 46.80 ,+37\ 48\ 33.1 & 18.93 & 51359 &  96 \\
{\it 73}  & 14\ 25\ 47.80 ,+37\ 47\ 48.0 & 18.70 & 52710 & 226 \\
{\it 74}  & 14\ 25\ 47.16 ,+37\ 47\ 32.7 & 19.84 & 50831 & 102 \\
{\it 75}  & 14\ 25\ 45.39 ,+37\ 46\ 38.2 & 18.75 & 51263 &  72 \\
{\it 76}  & 14\ 25\ 32.10 ,+37\ 49\ 42.1 & 19.94 & 52009 & 122 \\
{\it 77}  & 14\ 25\ 35.57 ,+37\ 47\ 55.9 & 18.80 & 51456 & 156 \\
     78   & 14\ 25\ 33.46 ,+37\ 47\ 57.8 & 21.05 & 80056 & 101 \\
   
            \noalign{\smallskip}			    
            \hline					    
            \noalign{\smallskip}			    
            \hline					    
         \end{array}
     $$ 
         \end{table}
\addtocounter{table}{-1}
\begin{table}[!ht]
          \caption[ ]{Continued.}
     $$ 
           \begin{array}{l c r r r}
            \hline
            \noalign{\smallskip}
            \hline
            \noalign{\smallskip}

\mathrm{ID} & \alpha,\delta\,(\mathrm{J}2000) & r^{\mathrm{Mega}} & v\,\,\,\,\,& \Delta v\\
 &                  & &\mathrm{(\,km}&\mathrm{s^{-1}\,)}\\

            \hline
            \noalign{\smallskip}
   
79        & 14\ 25\ 30.98 ,+37\ 47\ 42.2 & 19.72 &107174 &  52 \\
80        & 14\ 25\ 34.12 ,+37\ 46\ 19.4 & 19.63 & 63240 & 146 \\
81        & 14\ 25\ 33.41 ,+37\ 46\ 12.7 & 18.57 & 45921 & 120 \\
{\it 82}  & 14\ 25\ 40.35 ,+37\ 50\ 40.8 & 17.95 & 49856 &  26 \\
83        & 14\ 25\ 41.46 ,+37\ 50\ 37.9 & 19.52 & 47584 & 100 \\
84        & 14\ 25\ 30.96 ,+37\ 48\ 56.2 & 20.48 & 83757 & 122 \\
{\it 85}  & 14\ 25\ 42.09 ,+37\ 49\ 49.6 & 19.56 & 51428 &  78 \\
{\it 86}  & 14\ 25\ 44.52 ,+37\ 49\ 60.0 & 18.17 & 50404 &  60 \\
{\it 87}  & 14\ 25\ 41.35 ,+37\ 48\ 53.7 & 21.00 & 49505 & 268 \\
{\it 88}  & 14\ 25\ 50.54 ,+37\ 49\ 58.0 & 18.46 & 50589 &  76 \\
89        & 14\ 25\ 36.49 ,+37\ 47\ 37.6 & 18.77 & 55764 & 140 \\
{\it 90}  & 14\ 25\ 57.98 ,+37\ 50\ 23.9 & 19.99 & 53193 & 126 \\
91        & 14\ 25\ 49.31 ,+37\ 48\ 58.9 & 20.49 & 63713 & 212 \\
92        & 14\ 25\ 56.55 ,+37\ 49\ 45.4 & 17.43 & 55190 &  76 \\
{\it 93}  & 14\ 26\ 00.35 ,+37\ 50\ 04.3 & 17.88 & 48048 &  74 \\
94        & 14\ 25\ 47.90 ,+37\ 48\ 00.0 & 16.68 & 39699 &  60 \\
95        & 14\ 25\ 47.86 ,+37\ 47\ 40.3 & 19.48 & 55602 & 110 \\
{\it 96}  & 14\ 25\ 48.93 ,+37\ 47\ 40.5 & 18.77 & 50226 & 244 \\
{\it 97}  & 14\ 25\ 53.51 ,+37\ 48\ 05.9 & 17.88 & 46140 & 150 \\
98        & 14\ 25\ 56.78 ,+37\ 47\ 54.6 & 18.98 & 64149 & 234 \\
{\it 99}  & 14\ 25\ 42.34 ,+37\ 45\ 37.4 & 20.02 & 52653 & 100 \\
100       & 14\ 25\ 56.75 ,+37\ 47\ 27.0 & 19.59 & 55676 & 100 \\
{\it 101} & 14\ 25\ 54.29 ,+37\ 46\ 51.5 & 20.42 & 53514 & 126 \\
102       & 14\ 25\ 52.73 ,+37\ 45\ 45.9 & 18.49 & 62209 &  44 \\
{\it 103} & 14\ 25\ 49.27 ,+37\ 45\ 04.7 & 18.83 & 48960 &  80 \\
104       & 14\ 25\ 52.09 ,+37\ 45\ 16.2 & 19.67 &145911 & 156 \\
{\it 105} & 14\ 26\ 09.38 ,+37\ 47\ 29.2 & 18.65 & 52566 &  46 \\
{\it 106} & 14\ 25\ 49.55 ,+37\ 44\ 32.2 & 19.19 & 50106 &  46 \\
{\it 107} & 14\ 26\ 01.01 ,+37\ 45\ 49.6 & 18.61 & 47775 & 100 \\
{\it 108} & 14\ 26\ 07.52 ,+37\ 46\ 31.9 & 19.20 & 50245 &  80 \\
109       & 14\ 25\ 51.38 ,+37\ 43\ 58.6 & 19.98 & 82764 & 132 \\
110       & 14\ 25\ 51.90 ,+37\ 43\ 49.7 & 19.43 & 58839 & 152 \\
111       & 14\ 25\ 57.45 ,+37\ 44\ 19.9 & 19.63 & 63715 & 164 \\
112       & 14\ 25\ 50.55 ,+37\ 43\ 08.8 & 19.36 & 24748 & 100 \\
{\it 113} & 14\ 25\ 59.89 ,+37\ 44\ 11.9 & 19.12 & 50613 &  60 \\
{\it 114} & 14\ 25\ 59.47 ,+37\ 50\ 54.1 & 17.33 & 49944 &  52 \\
{\it 115} & 14\ 26\ 06.58 ,+37\ 51\ 28.3 & 17.18 & 49226 &  59 \\
{\it 116} & 14\ 25\ 29.47 ,+37\ 47\ 00.1 & 17.30 & 49505 &  37 \\
{\it 117} & 14\ 26\ 19.25 ,+37\ 48\ 43.5 & 17.19 & 49969 &  51 \\
{\it 118} & 14\ 25\ 49.81 ,+37\ 53\ 04.0 & 17.66 & 50095 &  43 \\
{\it 119} & 14\ 26\ 22.97 ,+37\ 54\ 01.9 & 17.36 & 51262 &  44 \\

            \noalign{\smallskip}			    
            \hline					    
            \noalign{\smallskip}			    
            \hline					    
         \end{array}
     $$ 
\end{table}

The brightest cluster galaxy is ID.~29 ($r^{\rm Mega}=15.87$, hereafter
BCG1). It lies at the south-west and is non-dominant (in luminosity) 
in the cluster. In fact, there is a second brightest cluster galaxy at the
north-east (ID.~20, $r^{\rm Mega}=16.39$, hereafter BCG2). The two
galaxies are separated by $\sim 1.7$\arcmin, i.e. $\sim 0.3$ \h at the
cluster distance.

\section{Analysis of the optical data}
\label{anal}

\subsection{Cluster member selection}
\label{memb}

\begin{figure}
\centering
\resizebox{\hsize}{!}{\includegraphics{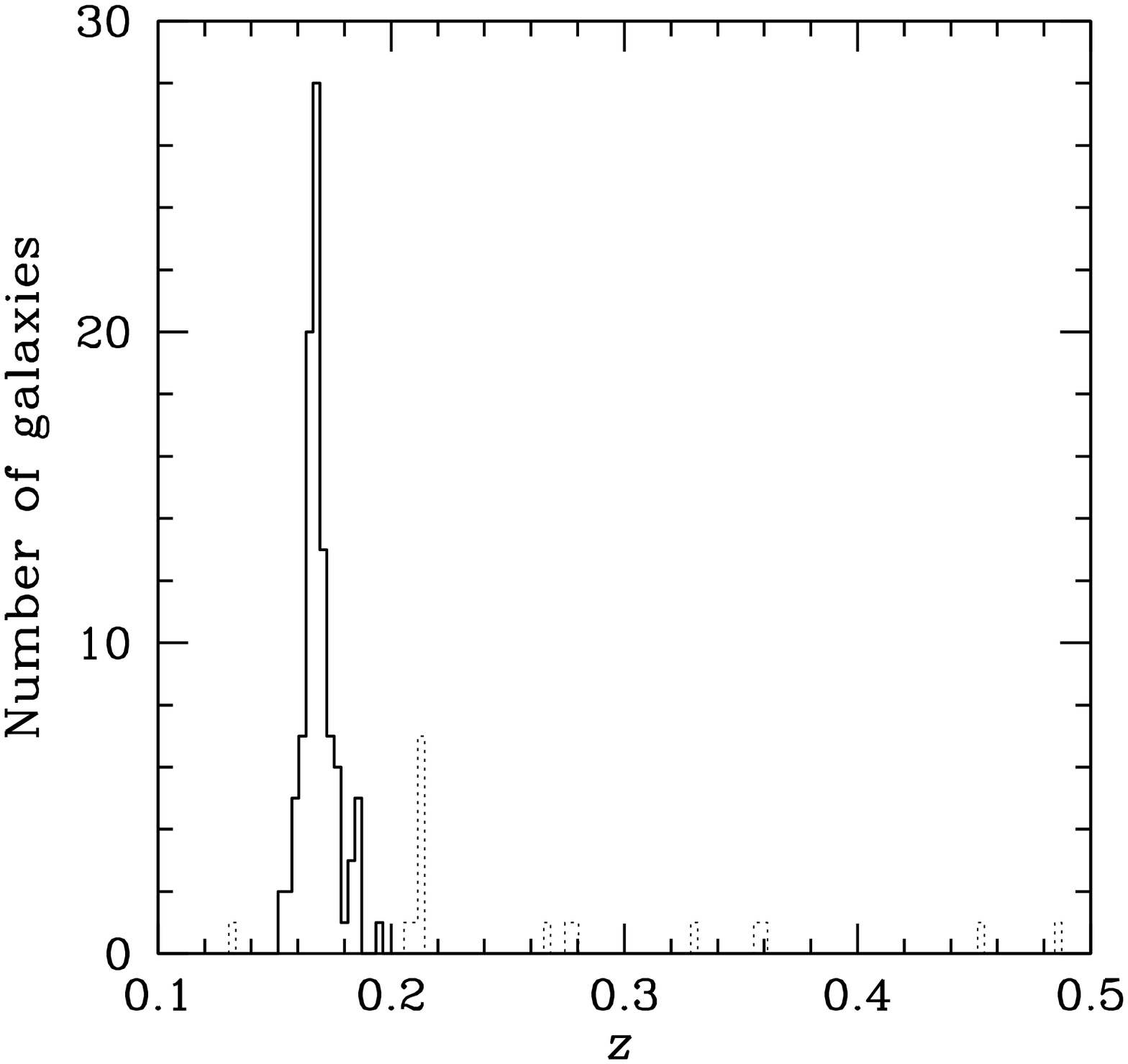}}
\caption{Redshift galaxy distribution. Solid line shows the histogram
corresponding to the 100 (provisional) galaxies assigned to
A1914 by the 1D-DEDICA reconstruction method.}
\label{fighisto}
\end{figure}

In order to select cluster members we followed a procedure with two
steps.  First, we run the 1D adaptive--kernel method (hereafter
1D-DEDICA, Pisani \citeyear{pis93} and \citeyear{pis96}; see also
Girardi et al.  \citeyear{gir96}; Fadda et al. \citeyear{fad96}). We
detected significant peaks (at $>$99\% c.l.) in the velocity
distribution. This procedure found A1914 as a peak at $z\sim0.1675$,
containing 100 (provisional) cluster candidates (in the range
$45\,921\leq v \leq 58\,839$ \kss, see Fig.~\ref{fighisto}).  We also
found 18 and 1 background and foreground galaxies, respectively.

\begin{figure}
\centering
\resizebox{\hsize}{!}{\includegraphics{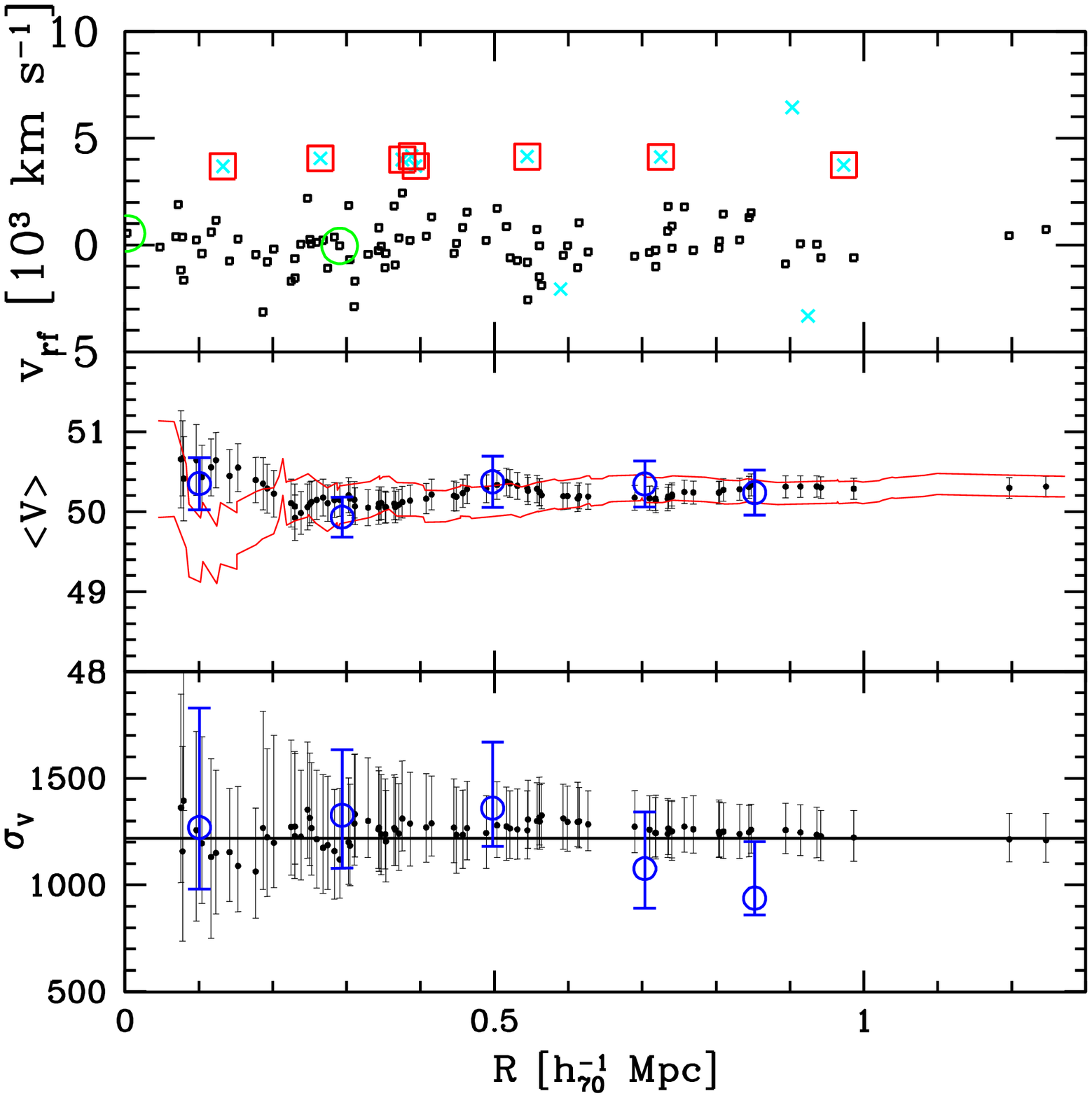}}
\caption
{{\em Top panel:} rest-frame velocity vs. projected distance (BCG1 is
assumed as the cluster center) for the all 100 galaxies within 
the velocity peak. Large green circles correspond to the location 
of the two brightest cluster galaxies. Cyan crosses show galaxies 
rejected as interlopers with the ``shifting gapper'' method. Galaxies 
belonging the HVG group are indicated with red squares. {\em Middle 
panel:} differential (blue circles) and integral (black dots) profiles 
of mean velocity for the 89 cluster members. We compute differential 
values within five annuli, each of $\sim 0.2$ \hh, from the cluster 
center. The integral profiles consider mean velocities of all galaxies 
within a given distance. The first value is obtained from the five inner
galaxy members. Error bars show uncertainties at $68\%$ c.l.. The
band within red lines indicates the integral mean velocity when
using BCG2 as the cluster center. {\em Bottom panel:} The same
as middle panel but for the $\sigma_{\rm V}$ profiles. The
horizontal line represents the value for the X-ray temperature (9
keV) transformed into $\sigma_{\rm V}$ assuming a $\beta_{\rm
spec}=1$ model for the density-energy distribution between ICM and
galaxies (see Sect.~\ref{disc}).}
\label{figprof}
\end{figure}

Then, in a second step, we only consider the 100 likely cluster
candidates to run the ``shifting gapper'' method proposed by Fadda et
al.  (\citeyear{fad96}; see also, e.g., Girardi et
al. \citeyear{gir11}), which takes into account a combination of
velocity and position of the galaxies. This method needs the
definition of a cluster center, but the optical center of A1914
is not obvious due to the absence of a clear dominant galaxy and to
the offset of the X-ray center (e.g., Maughan et
al. \citeyear{mau08}). So, we decided to assume as cluster center the
position of BCG1 (see Table~\ref{catalogA1914}). The
application of the ``shifting gapper'' rejected another eleven
galaxies leading to a final sample of 89 cluster members
(Fig.~\ref{figprof} -- top panel).

In order to check the robustness in the galaxy member selection 
and estimate how the choice of cluster center could affect this
selection, we executed the shifting gapper method, but
now considering the BCG2 as cluster center. This procedure selected 
identical galaxy members. That is, in our case, the galaxy member 
selection, and so the dynamical analysis here exposed is not affected 
by the choice of the cluster center. So, we decided to consider the 
BCG1 as cluster center, as in agreement with Okabe \& Umetsu 
(\citeyear{oka08}).

\subsection{Global cluster properties}
\label{glob}

By using the biweight method (Beers et al. \citeyear{bee90}, ROSTAT
software) with the 89 cluster members, we obtained a mean cluster
redshift of $\left<z\right>=0.1678\pm 0.0004$, i.e. $\left<v\right>=
(50\,313\pm$140) \kss. In addition, by applying the same method and
correcting for cosmological effects and standard velocity errors
\citep{dan80}, we obtained $\sigma_{\rm V}=1210_{-110}^{+125}$ \kss
(errors were estimated using the bootstrap technique). However, in
order to check the robustness of this estimate, we study the
variation of $\sigma_{\rm V}$ with the distance to the cluster center
(Fig.~\ref{figprof}, bottom panel). The integral $\sigma_{\rm V}$
profile is flat, suggesting that the estimation of $\sigma_{\rm V}$ is
robust. Furthermore, when considering members within 0.1 \h from 
the BCGs (that is two groups of 7 and 8 galaxy members around BCG1 and 
BCG2, respectively), we measure similar mean velocities. We only find 
a modest difference, being $\left<v\right>_{\rm BCG2}<\left<v\right>_{\rm BCG1}$,
which is in agreement with the finding that $v_{\rm BCG2}<v_{\rm BCG1}$
(see also section \ref{specphoto}). 

\subsection{The small high velocity group}
\label{HVG}

Figure~\ref{figprof} (top panel) shows that most of the
interlopers have a very similar high velocity.  We assign eight
galaxies to a likely galaxy group (red squares in Fig.~\ref{figprof},
top panel). Seven out of these eight galaxies lie in the southwest
cluster region, thus reinforcing the idea that this is a real
structure. For this high velocity galaxy group (hereafter HVG) we
estimate $\left<v\right>_{\rm HVG}=(55\,557\pm$89) \ks and
$\sigma_{{\rm V}, \rm{HVG}}=221_{-46}^{+55}$ \kss.

\subsection{Velocity distribution and 3D substructure}
\label{velo}

Deviations from Gaussianity in the velocity distribution are
interpreted as an important sign that clusters present a complex
dynamics \citep{rib11}.

In order to check the Gaussianity in the velocity distribution,
we used three profile estimators. These are the skewness, the
kurtosis, and the scaled tail index STI (see Bird \& Beers
\citeyear{bir93}).  The STI finds evidence for non-Gaussianity
at about $95\%$-$99\%$ c.l., suggesting a heavy tailed distribution
(see Bird \& Beers \citeyear{bir93} and their Table~2).

\begin{figure}
\centering 
\resizebox{\hsize}{!}{\includegraphics{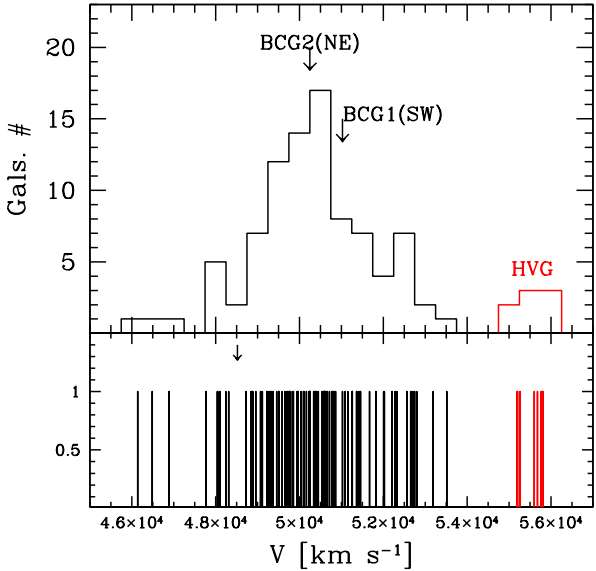}}
\caption{{\em Upper panel}: Velocity distribution of the galaxy
  members (black line) and to the HVG (red line). Velocities of 
  BCG1 and BCG2 are also indicated by arrows. {\em Lower panel}:
  Streak density diagram of cluster members (black) and HVG members
  (red). The significant gap in the velocity distribution is indicated
  by an arrow.}
\label{figstrip}
\end{figure}

Furthermore, we investigated the presence of gaps in the velocity
distribution. By using the weighted gap analysis presented by Beers et
al. (\citeyear{bee91}; \citeyear{bee92}; ROSTAT software), we detect
one significant gap at the $97\%$ c.l.. This gap divides A1914 into
two groups, composed of 9 and 80 galaxies at low and high
velocities, respectively. By applying the 2D Kolmogorov-Smirnov
test \citep{fas87} we see that the galaxies of the two groups
present the same spatial distribution (see Fig.~\ref{figstrip}).

We also applied the 1D-Kaye's mixture model (Ashman et
al. \citeyear{ash94}; see also, e.g., Boschin et al. \citeyear{bos12})
-- hereafter 1D-KMM --  to search for bimodal partitions
significantly fitting to the velocity distribution. The most
likely solution (well under the $90\%$ significance c.l.) indicates
two groups of 15 and 74 galaxies, spatially not differing. Considering
the KMM results, which take into account the group membership
probability, we obtain that the two groups differ for about $\sim 140$
\ks in the cluster rest-frame ($\left<v\right> _{\rm
KMM1D-LV}=50\,155$ and $\left<v\right>_{\rm KMM1D-HV}=50\,319$
\kss) and the high velocity group has a much higher velocity
dispersion ($\sigma_{{\rm V},{\rm KMM1D-HV}}\sim 980$ \ks 
vs. $\sigma_{{\rm V}, {\rm KMM2D-LV}}\sim 330$ \kss).

\begin{figure}
\centering
\resizebox{\hsize}{!}{\includegraphics{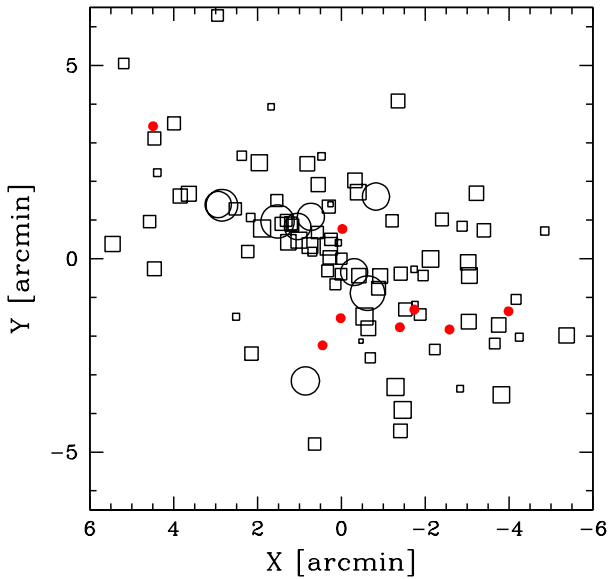}}
\caption
{Projected spatial distribution of 89 cluster members (open symbols)
and the 8 members of HVG (red solid circles). The size of symbols is
directly correlated with the value of the radial velocity for
each galaxy. Circles and squares indicate the two groups at low
and high velocities, as detected throught the weighted gap
analysis. The plot is centered on the BCG1 position.}
\label{figgrad}
\end{figure}

Correlations between spatial and velocity distributions of cluster
galaxies usually indicate the presence of actual substructures.
With this idea in mind, we used several techniques to reveal the
structure of A1914 by combining positions and velocities. First, we
searched for velocity gradients in the plane of the sky by performing
multiple linear fits. The results of this test reveals no evidences of
gradients. In addition, we performed a set of 3D tests: the classical
$\Delta$ statistics \citep{dre88}, as well as its variation
which considers separately mean velocity and velocity dispersion
kinematical indicators (Girardi et al. \citeyear{gir97}; Ferrari et
al. \citeyear{fer03}); the $\alpha$-test \citep{wes90} and the
$\epsilon$-test \citep{bir94} based on the projected mass predictions.
None of the above mentioned tests yielded positive detection of
substructures.  Moreover, we found no substructure by applying the
technique developed by \cite{ser96}, also named the ''Htree-method''
(see also, Durret et al.  \citeyear{dur10}; Boschin et
  al. \citeyear{bos12}).

\subsection{2D galaxy distribution of the spectroscopic catalog}
\label{specphoto}

\begin{figure}
\includegraphics[width=8cm]{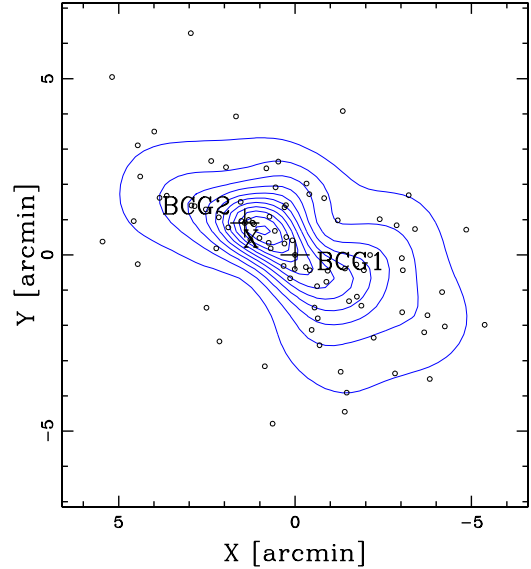}
\caption{Projected spatial distribution with relative isodensity
contours of spectroscopic cluster members (small circles). The
density map has been obtained with the 2D-DEDICA method. BCG1
position is assumed as the cluster center. Both BCG1 and BCG2
locations are shown by crosses. The ``X'' indicates the X-ray peak
\citep{gov04}.}
\label{figk2z}
\end{figure}

We applied the 2D adaptive-kernel technique (hereafter
2D-DEDICA) to the spatial distribution of member galaxies. This
method found only one significant peak, lying close to BCG2
and elongated toward BCG1. Figure~\ref{figk2z} shows that A1914
presents an elongated profile in the NE-SW direction, suggesting a
bimodal structure. To further investigate this point we applied
the 2D-KMM method using as seeds the cluster members contained
within 0.1 \h from BCG1 and BCG2 (7 and 8 galaxies,
respectively). The method detects a bimodal solution, significant at
the $99.6\%$ c.l., with the KMM2D-SW group of 65 galaxies at
south-west and the KMM2D-NE group of 24 galaxies at north-east (but
note that both BCG1 and BCG2 are now both assigned to the SW
group). When applying the 3D-KMM method we still found a bimodal
solution with two groups of 69 (SW) and 20 galaxies (NE), with BCG1
and BCG2 assigned to the SW and NE groups, respectively. However, the
significance of the 3D result is only at the $96\%$ c.l., thus
indicating that velocities do not give a positive contribution to the
separation of the two substructures. In fact, according to the
Kolmogorov-Smirnov test, there is no difference between the
velocity distributions of the two 3D groups.

\subsection{2D galaxy distribution of the photometric catalogs}
\label{photo}

Our spectroscopic sample does not map the whole cluster field, besides
of suffering for magnitude incompleteness. To overcome these
restrictions we resorted to the photometric catalogs.

\begin{figure}
\centering
\resizebox{\hsize}{!}{\includegraphics{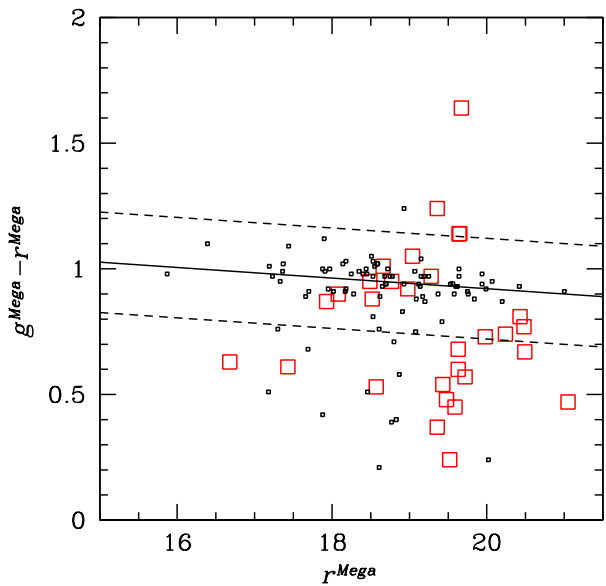}}
\caption
{CFHT $g^{\rm Mega}-r^{\rm Mega}$ vs. $r^{\rm Mega}$ diagram for 
galaxies with available spectroscopy. Black and red squares indicate 
member and non-member galaxies. The solid line gives the CMR determined 
on member galaxies; the dashed lines are drawn at $\pm$0.2 mag from 
this value.}
\label{figcm}
\end{figure}

Using the CFHT photometry, we construct ($g^{\prime}$--$r^{\prime}$ 
vs. $r^{\prime}$) color-magnitude relation (hereafter CMR, see
Fig.~\ref{figcm}), and select likely early-type members within the red 
sequence (RS) locus. In order to compute the RS, we apply a 2$\sigma$-clipping 
fitting procedure to the spectroscopic cluster galaxies. We obtained 
$g^{\rm Mega}$--$r^{\rm Mega}$=1.341-0.021$\times r^{\rm Mega}$ on 63 
spectroscopic cluster members. With this method we selected as likely 
cluster members the objects within $\pm0.2$ mag with respect to the RS. 
Figure~\ref{figcm} shows as the selected magnitude 
intervals seem adequate to select RS galaxies. In this way, we only use 
good tracers of the cluster galaxy population (e.g., Lubin et al. 
\citeyear{lub00}). However, Fig.~\ref{figcm} suggests that our selection 
of photometric members is somewhat contaminated by several non members, 
probably because A1914 is contained in a rich large scale structure (see 
in the following).

\begin{figure}
\centering
\includegraphics[width=6.6cm]{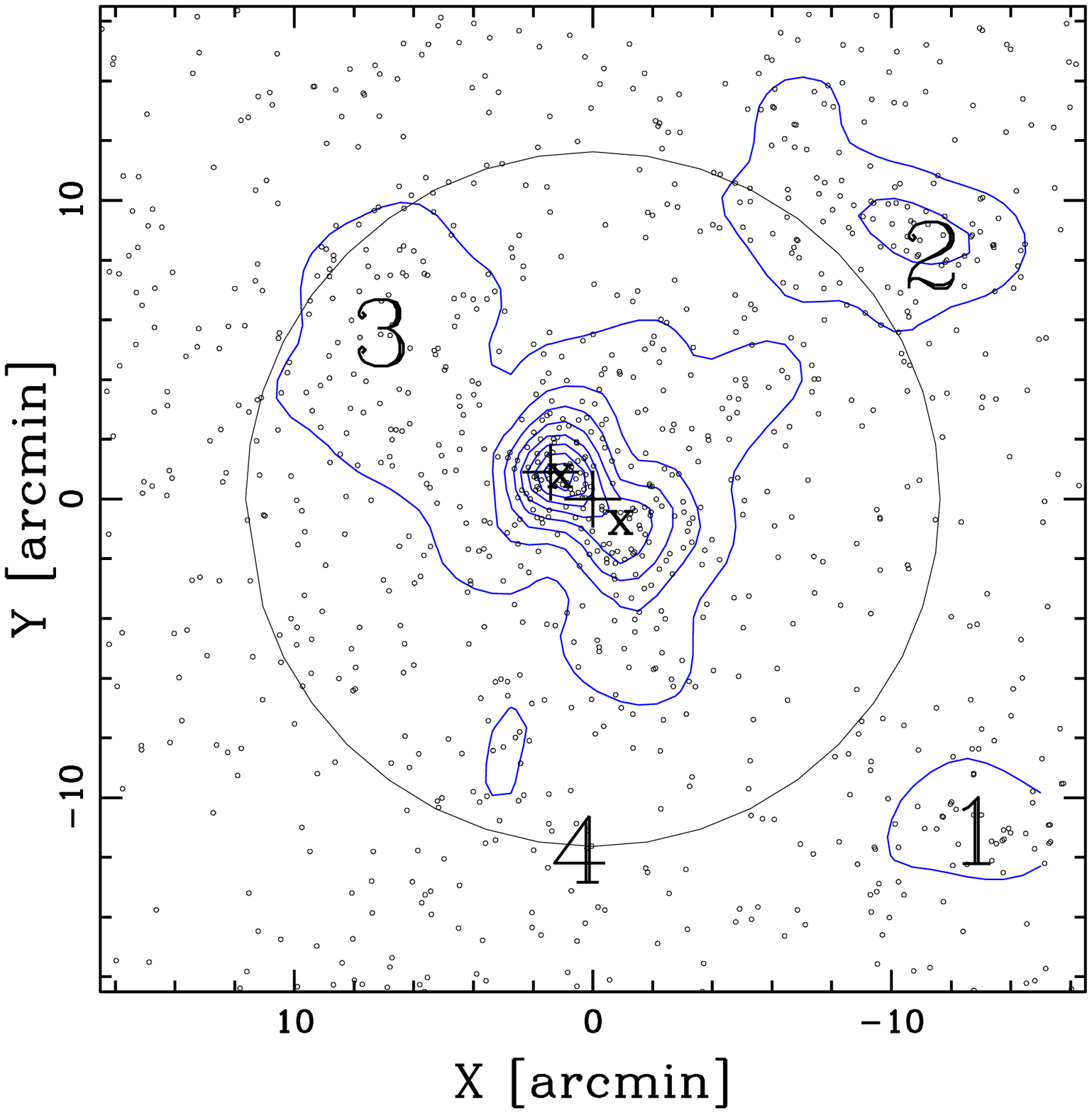}
\includegraphics[width=6.6cm]{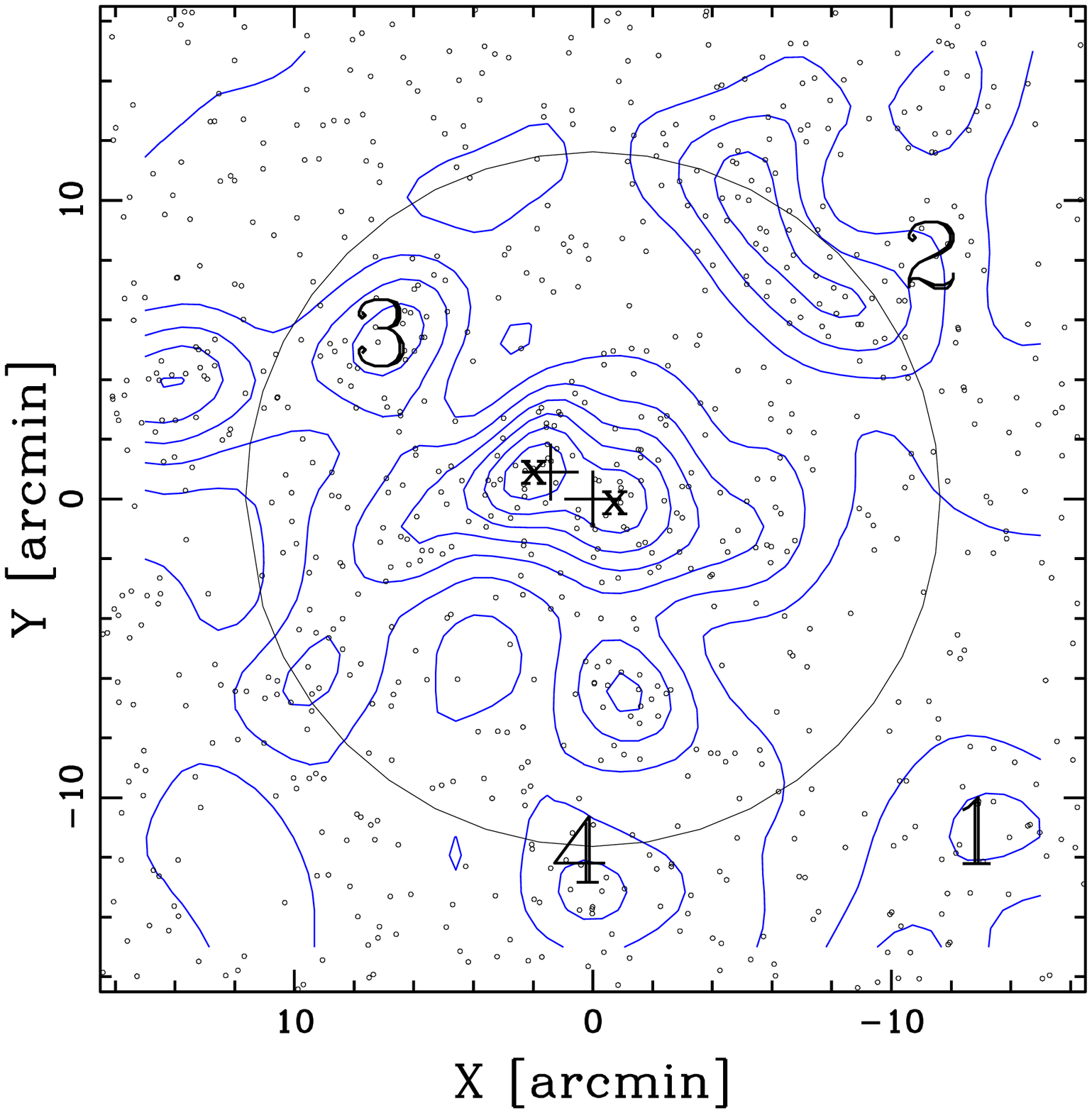}
\includegraphics[width=6.6cm]{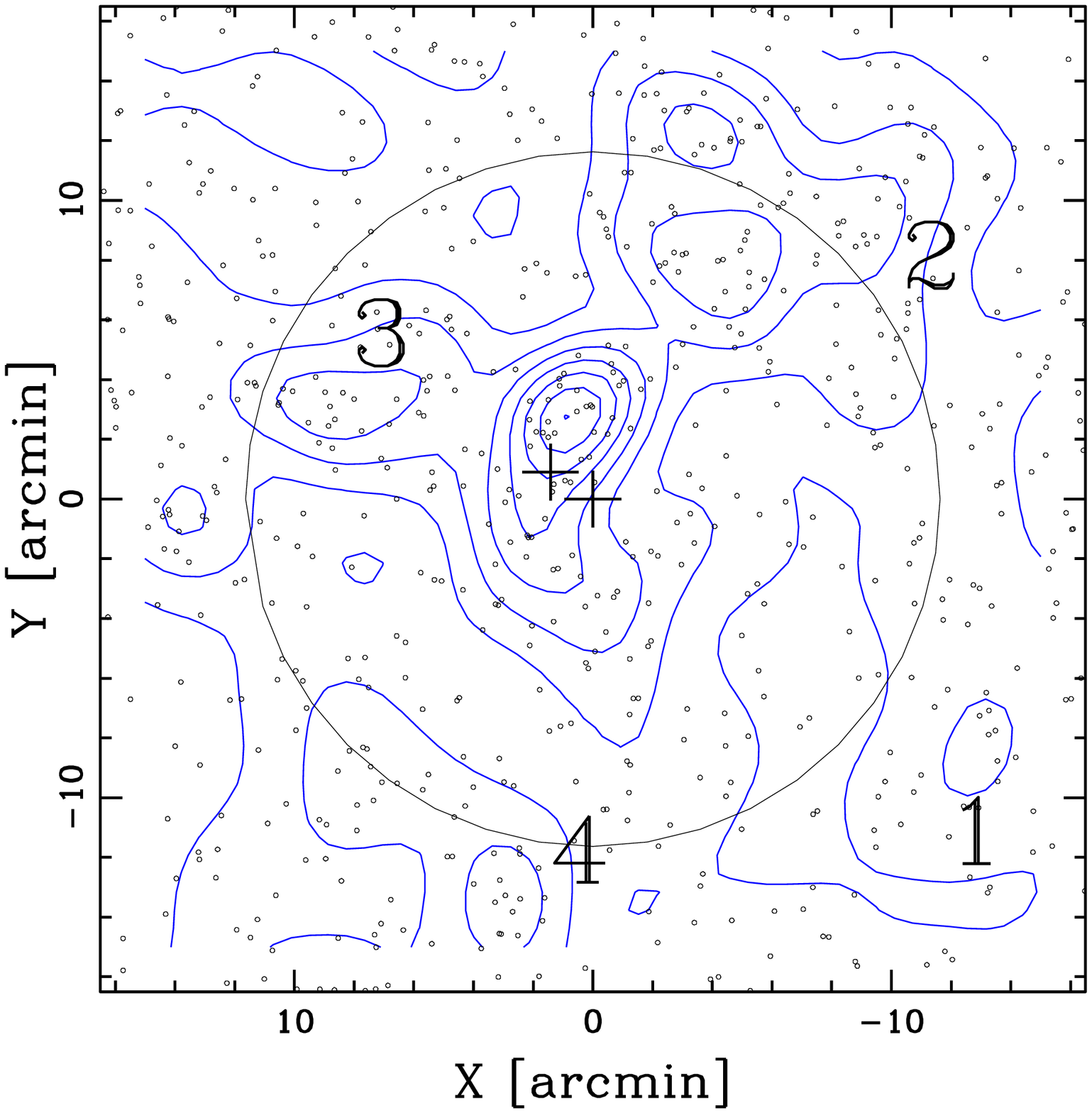}
\caption{Projected spatial distribution and isodensity contours (CFHT likely 
galaxy members with $r^{\rm Mega}<21$ ({\em Top panel}), 
$21\le r^{\rm Mega}< 22$ ({\em Middle panel}), and $22\le r^{\rm Mega}< 22.5$ 
({\em Bottom panel}). Contours were estimated using the 2D-DEDICA method. 
"$\times$" symbols note the NE and SW clumps in the upper and middle panel, 
while "$+$" symbols mark the position of BCG1 and BCG2. The numbers indicate 
the positions of the four galaxy clusters listed by NED 
within 20\arcm from the cluster center and having estimated or photometric 
redshift close to A1914 ($\Delta z<0.06$), in order of increasing $\Delta z$. The 
circles indicate the cluster region within 2 \h at the cluster distance, i.e. 
somewhat smaller than the virial radius (2.2-2.7 \h, which is determined by 
the adopted model).}
\label{figk2}
\end{figure}

Figure~\ref{figk2} shows the isodensity contours of likely cluster
galaxies in three different magnitude bins considering the CFHT
photometry.  The three bins contain a comparable number of galaxies
(from 2065 to 2377 in the whole CFHT area). A1914 is well detected
using the most luminous galaxies and, as expected due to contamination
problems, the noise increases at fainter magnitudes. The results
for $r^{\rm Mega}> 22$ are likely very contaminated by background
systems (see below) and are shown for comparison. Through the
photometric sample we can assess that the cluster structure is really
NE-SW elongated and that the highest peak is not centered on
BCG1. The precise position of the highest peak depends on the
magnitude bin or on the catalog, however it is close to BCG2. In 
addition, using CFHT data we found an important secondary peak lying at
south west of BCG1. In the $r^{\prime}< 21$ sample, the two peaks
are separated by $\sim 2.5$\arcmin, i.e. $\sim 0.4$ \h at the cluster
distance.

We also note that the external regions of A1914 are
particularly rich of structures. Figure~\ref{figk2} shows the position
of four galaxy clusters listed by NED: 1. NSCS J142452+373753 at
$z\sim 0.17$ (identified in the Digitized Second Palomar Observatory
Sky Survey DPOSS, Lopes et al. \citeyear{lop04}; and then in SDSS);
2. 400d J1425+3758 at $z\sim 0.17$ (identified in the 400 Square
Degree ROSAT PSPC Galaxy Cluster Survey, Burenin et
al. \citeyear{bur07}); 3.  NSCS J142638+375327 at $z\sim 0.20$; 4.
GMBCG J216.49530+37.62102 at $z\sim 0.23$ (identified in the SDSS DR7,
Hao et al. \citeyear{hao10}).  When using the best sample (the one
with $r^{\rm Mega}<21$), the two clusters closest in redshift (N.1 and
N.2) are well detected, while the clusters N.3 and N.4 are better
detected in the deeper magnitude ranges, thus confirming us that we
are including more and more interlopers among our likely cluster
members when considering fainter galaxies.
 
Table~\ref{tabdedica2d} lists information for the four highest,
significant peaks in the galaxy distribution using the CFHT data: the
estimated number of likely members, $N_{\rm S}$ (Col.~2); Equatorial
coordinates of the substructure (Col.~3); the relative isodensity
respect the highest peak, $\rho_{\rm S}$ (Col.~4); the $\chi^2$ for
each clump (Col.~5). Galaxy clusters No.1 and No.4 are detected as
significant peaks in the $21 \le r^{\rm Mega}<22$ sample, too, but having 
lower density.

\begin{table}
        \caption[]{2D substructure detected in the CFHT photometric data.}
         \label{tabdedica2d}
            $$
         \begin{array}{l r c c c }
            \hline
            \noalign{\smallskip}
            \hline
            \noalign{\smallskip}
\mathrm{2D-Subclump} & N_{\rm S} & \alpha,\,\delta\,({\rm J}2000)&\rho_{\rm S}&\chi^2_{\rm S}\\
(r=r^{\rm Mega}) & & \mathrm{14:m:s,+37:\arcmm:\arcs}&&\\
         \hline
         \noalign{\smallskip}
\mathrm{NE}\ (r<21)  	   &167&26:02.2,\ 49:46&1.00&104\\
\mathrm{SW}\ (r<21)  	   &176&25:52.0,\ 48:12&0.54&53\\
\mathrm{No.2}\ (r<21)	   &204&25:02.2,\ 57:57&0.27&44\\
\mathrm{No.1}\ (r<21)	   & 99&24:49.4,\ 37:56&0.20&26\\
\mathrm{NE}\   (21\le r<22)     & 90&26:06.7,\ 49:53&1.00&24\\
\mathrm{SW}\   (21\le r<22)     & 96&25:53.0,\ 48:52&0.86&18\\
\mathrm{No.3}\    (21\le r<22)  & 46&26:30.8,\ 54:25&0.62&15\\
\mathrm{No.2}\    (21\le r<22)  & 80&25:30.4,\ 59:13&0.62&13\\
              \noalign{\smallskip}
              \noalign{\smallskip}
            \hline
            \noalign{\smallskip}
            \hline
\small
         \end{array}
$$
         \end{table}

\section{Discussion}
\label{disc}

We estimate a high value of the velocity dispersion, $\sigma_{\rm
  V}=1210_{-110}^{+125}$ \kss. This result well agrees with a hot ICM
showing a mean $T_{\rm X}\sim 9$ keV (Baldi et al. \citeyear{bal07}
and Maughan et al. \citeyear{mau08}) when assuming energy
equipartition between galaxies and gas energy per unit
mass\footnote{$\beta_{\rm spec}=\sigma_{\rm V}^2/(kT_{\rm X}/\mu
  m_{\rm p})$ with $\mu=0.58$ the mean molecular weight and $m_{\rm
    p}$ the proton mass.}, i.e. $\beta_{\rm spec}=1$, both suggesting
a massive galaxy cluster. In the following sections we discuss 
our findings on the dynamical mass and cluster structure. Thus, based on 
these results, we propose a two-body model and time scale for the 
collision of substructures in Abell 1914.

\subsection{Mass estimates}
\label{mass}

We computed the global virial quantities assuming the dynamical
equilibrium (but see in the following) and in the framework of usual
assumptions, i.e. cluster sphericity and coincidence in the
galaxy-mass distributions. Following the method detailed in Girardi \&
Mezzetti (\citeyear{gir01}, see also Girardi et al. \citeyear{gir98})
we obtained $R_{\rm vir}$, an estimation of $R_{\rm 200}$, and the
mass within this radius. We assume a quasi-virialized region of
$R_{\rm vir}=0.17\times \sigma_{\rm V}/H(z)$ \h (see Eq.~1 of Girardi
\& Mezzetti \citeyear{gir01} with the corresponding scaling of $H(z)$
from Eq.~ 8 of Carlberg et al. \citeyear{car97} for $R_{200}$). We
estimate the mass using the equation $M=M_{\rm vir}-SPT=3\pi/2 \cdot
\sigma_{\rm V}^2 R_{\rm PV}/G-SPT$ (Eq.~3 of Girardi \& Mezzetti
\citeyear{gir01}, with ${\rm R}_{\rm PV}$ derived as described
in Eq.~13 of Girardi et al. (\citeyear{gir98}), with $A=R_{\rm vir}$
as a close approach, and being the surface pressure term correction
(SPT) a 20\% of $M_{\rm vir}$. Both $R_{\rm vir}$ and $M$ were
estimated considering the velocity dispersion with the usual
scaling-laws, where $R_{\rm vir} \propto \sigma_{\rm V}$ and
$M(<R_{\rm vir}) \propto \sigma_{\rm V}^3$. We obtained $M(<R_{\rm
  vir}=2.7 \hhh)=2.6\pm 0.8$\mquii.

On the other hand, it is commonly accepted that A1914 hosts a 
merging event, and so velocity dispersion and X-ray temperature 
could be enhanced (e.g., Ricker \& Sarazin \citeyear{ric01};
Schindler \& M\"uller \citeyear{sch93}). Our analysis fails to
separate the cluster substructures in the velocity space. We have to
assume the result derived from the 1D-KMM method (Sect.~\ref{velo},
i.e. the two best Gaussians obtained there, despite their low
significance. According to these results, the secondary group presents
a very small velocity dispersion and its mass can be neglected with
respect to the main system. For the main system $\sigma_{\rm
V,main}\sim 980$ \ks which leads to $M(<R_{\rm vir}=2.2 \hhh)=1.4$
\mquii. Hereafter, we consider as reliable the mass range $M_{\rm
sys}=1.4-2.6$ \mqui for the whole A1914 system.

The above value of $\sigma_{\rm V,main}$ is in agreement with
$\sigma_{\rm V,SIS}=846\pm87$ estimated from the weak lensing analysis
by \citet{oka08}. For a punctual comparison with the projected mass
computed by \citet{oka08} within R=7 \arcmin, we project and
rescale our mass estimate $M_{\rm sys}$ assuming the cluster follows
a NFW profile, taking a mass concentration parameter $c$ from
\citet{nav97} and correcting by the factor $1+z$ (Bullock et
al. \citeyear{bul01}; Dolag et al. \citeyear{dol04}, here $c\sim
4$). We obtain $M_{\rm sys,2D}(<R=1.2 \hhh) =(1.1$--$2.0)$\mquii,
in agreement with that $M_{\rm 2D}(<R=7 \arcmin)=(4.10\pm 1.55)$
\mqua is a lower bound to the true enclosed mass \citep{oka08}.
Instead, there is some tension between our value of $\sigma_{\rm
V,main}$ and the velocity dispersion estimate computed using
redshifts from the on-going Hectospec Cluster Survey \citep{rin10},
$\sigma_{\rm V,Rines}= 698_{-38}^{+46}$ \kss. This explains the
difference between our and their virial mass estimate $M_{\rm
200,Rines}\sim 6.0$ \mquaa, as obtained rescaling the value of
$M_{100}$ ($R_{100}\sim 1.3 R_{200}$ and thus
$M_{200}\sim 0.9 M_{100}$ for a NFW profile \citep{eke96}.

\subsection{Cluster structure}
\label{stru}

Our 2D analyses confirm the existence of an important bimodal
structure elongated in the NE-SW direction and find two significant
peaks: the NE one closer to BCG2 and the SW one closer to 
BCG1, although not perfectly centered on the two BCGs.

The analysis of \cite{gov04} shows that the X-ray peak is displaced 
with respect to the BCGs positions. They find that
X-ray maximum is at South with respect to BCG2 (see their Fig.~5a).
 Similarly, we also find that the X-ray peak does not coincide 
with our peaks in the galaxy density (see our Fig.~\ref{figimage}). The 
offset between the optical and X-ray peaks suggests a post-merger cluster 
but, as noted by \cite{gov04}, the X-ray features are not typical (e.g., 
simulations by Roettiger et al. \citeyear{roe98}). In fact, their 
analysis shows the presence of a NE-SW arclike hot region crossing 
through the cluster center, while the X-ray emission is elongated in 
the WNW-ESE direction, someway perpendicular to that described by the two
galaxy/mass concentrations. Govoni et al. (\citeyear{gov04}) interpreted
this observational scenario as due to a large impact parameter merger.

We add four contributes to the comprehension of the A1914 merging
scenario.  These contributes are: 1) the relative importance of
the NE and SW subclusters; 2) the merger is likely most contained
in the plane of the sky, as suggested by the failure of 1D and
3D substructure analysis methods in detecting the two galaxy subclumps
(Pinkney et al. \citeyear{pin96}); 3) A1914 is embedded in a rich
large scale structure that suggests that cluster accretion
happens along two specific directions, NE-SW and NW-SE; 4) the
presence of HGV, a minor external group of uncertain nature.

Point 1. The NE subcluster, close to BCG2 and the X-ray peak, has
higher density than the SW subcluster, close to BCG1. However,
the SW subcluster is the richer - and likely the more massive as
shown by the respective galaxy population at end of
Sect.~\ref{specphoto} and Table~\ref{tabdedica2d}, and by the
respective velocity dispersions).  This agrees with the result of
the gravitational lensing analysis by Okabe \& Umetsu
(\citeyear{oka08}), where the peak C1, which is related to BCG2, is
the highest density peak in the mass distribution, while the second
density peak, C2 related to BCG1, is suggested to be the primary
cluster center.

The point 2. suggests a similarity between A1914 and the well-known
cluster Abell 754 (hereafter A754). A754 is a bimodal cluster where
two obvious substructures collide in the plane of the sky, showing an
X-ray emission profile elongated and perpendicular to that of the two
optical clumps \citep{zab95}. The X-ray peak is closer to the
denser optical peak, reminding us once more the similarity with
A1914. As for A754, the gross X-ray morphology has been firstly
explained with a non-zero impact parameter \citep{hen95}, but more
recent and detailed analyses suggest a more complex merger scenario,
possibly involving a third substructure or a mass of cool gas
disengaged from its previous host galaxy group and presently sloshing
\citep{mar03}. In fact, a shock is found in front of the denser
optical peak, as usual in head-on mergers \citep{mac11}.

Point 3. According to their estimated redshift, both galaxy
systems No.1 and 2 are likely to be connected to A1914.  No.1,
i.e. the cluster NSCS J142452+373753 \citep{lop04}, lies at SW.
This and the merger axis of the two subclusters strongly supports
the idea that A1914 is accreting groups from a filament aligned with
the NE-SW direction.  No.2, i.e. the cluster 400d J1425+3758
\citep{bur07}, is a massive system (with $T_{\rm X}> 5$ keV since
included in the 400 Square Degree ROSAT PSPC Galaxy Cluster Survey)
lying at NW, at the border of the virial radius of A1914, somewhat
along the direction of the NW-SE elongated bright feature of the
radio emission. Thus, No.2 traces the NW-SE direction of a likely
second filament accreting onto the cluster. This suggests a merging
scenario more complex than the simple bimodal one for A1914.

Point 4. The group HVG, although very poor, is also clearly
detected as a separate group in the projected phase space (see
Fig.~\ref{figprof}). Its nature is instead not clear. In this sense,
we think that the main collision between the two substructures may
have produced out-flying galaxies as predicted by simulations (e.g.,
Czoske et al. \citeyear{czo02}; Sales et al. \citeyear{sale07}) and 
detected in a few clusters (e.g., in Abell 3266 by Quintana et
al. \citeyear{qui96} and Flores et al. \citeyear{flo00}; in
Cl0024+1654 by Czoske et al. \citeyear{czo02}). This ipothesis is
suported by the fact that the system has not a circular morphology
but 6 out of 8 galaxies trace a large strip. Moreover, 4 out 8
galaxies show emission lines, thus suggesting a possible star
forming activity triggered by the cluster merger. Alternatively,
HVG could be a group in pre-collision phase with A1914, as well
as a completely unbound group. Anyway, it is likely so poorly massive 
that can be considered a secondary detail in the cluster dynamics.

\subsection{Bimodal model}
\label{bim}

In this section, we present our efforts to unravel the dynamics
of the merger between the two main subclusters in the SW-NE direction
with a simple bimodal model, assuming that this collision causes (at
least part of) the diffuse radio emission.  Following the method detailed for
other DARC clusters (see e.g., Abell 520 in Girardi et
al. \citeyear{gir08} and Abell 2345 in Boschin et al.
\citeyear{bos10}), we apply the two-body model \citep{bee82,tho82}
to evaluate the timescales of the merger. This simple model
assumes two point-mass bodies and a zero impact parameter.  The
model takes into account three parameters. These are the mass of the
whole system, $M_{\rm sys}$ (1.4--2.6 \mquii, see above), the relative
line of sight velocity in the rest-frame, $\Delta V$, and the
projected linear distance between the two substructures, $D$. As for
the relative motion parameters, we consider our more reliable
results, i.e. the estimate $\Delta V=140$ \ks obtained from our 1D
analysis, and the estimate $D\sim 0.4$ \h obtained from our 2D
analysis. Due to the small radiative life of relativistic electrons
and as a comparison with other radio halos clusters (e.g.,
Barrena et al. \citeyear{bar02}; Girardi et al. \citeyear{gir08}), we
assume an elapsed time, $t$, for the core crossing of few fractions of
Gyr. We consider two different cases: $t=0.1$ Gyr and $t=0.3$ Gyr.

\begin{figure}
\includegraphics[width=8cm]{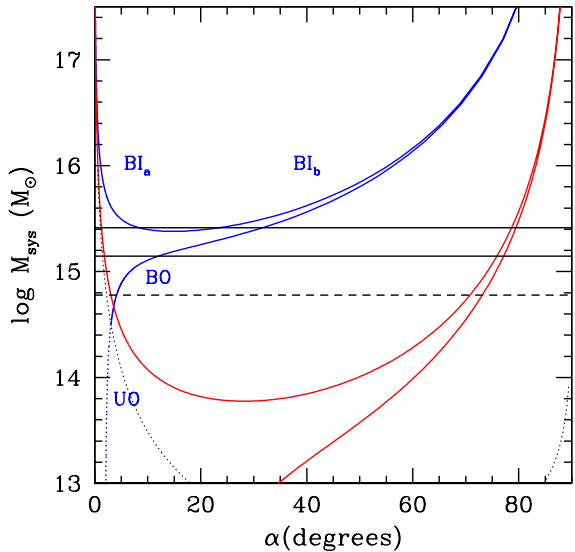}
\caption{Two--body model applied to the NE and SW galaxy subclusters.
The solutions are plotted as system mass vs. projection angle. Thick 
solid and thick dotted curves are bound and unbound solutions, respectively.
Blue and red lines correspond to the case of $t=0.1$ and $t=0.3$ Gyr, 
respectively. BI$_{\rm a}$ and BI$_{\rm b}$ labels refer to bound and 
incoming, which denote the collapsing solutions (solid curve). On the
other hand, the expanding solutions and unbound outgoing solutions 
(solid curve going on in the dotted curve, respectively) are labeled
as BO and UO. Labels for the $t=0.3$ case are skipped to keep the figure
clear. Our mass estimate is enclosed by the horizontal solid 
lines, while the dashed line is the mass value obtained by Rines et al.
(\citeyear{rin10}). The Newtonian criterion predicts a limit for bound 
solutions; this limit is shown as the thin dashed curve (above and below 
is the bound and unbound regimes, respectively).}
\label{figbim}
\end{figure}

Figure~\ref{figbim} compares the model solutions as a function of
$\alpha$, where $\alpha$ is the projection angle between the plane of
the sky and the line connecting the centers of the two clumps, 
with the mass estimate of the system $M_{\rm sys}$. At $t=0.1$
Gyr, the solution is bound and outgoing (BO) with $\alpha \sim
10\degree-25\degree$, in agreement with the fact that we expect
a merging axis mostly contained in the plane of the sky. At $t=0.3$
Gyr, the model predicts unlikely angles $\alpha>60\degree$.  Even
when considering as $M_{\rm sys}$ the virial mass value by Rines et
al. (\citeyear{rin10}, see our Sect.~\ref{mass}), the model with
$t\sim 0.1$ Gyr should be preferred.

\section{Conclusions}
\label{conc}

In conclusion, A1914 shows clear evidence of a recent cluster
merger along the NE-SW direction and almost contained in the plane 
of the sky. The presence of an ongoing merger and the large mass are 
the main features of typical clusters with radio halos described in 
the literature. The large scale structure in the environment of
A1914 suggests evidence of a sencond direction of cluster 
accretion, NW-SE. This merging axis is likely related to the bright 
feature of the diffuse radio emission. Thus, we argue that the
unusual radio appearance of A1914 is due to the complexity of the
merger. Indeeed, we point out that A1914 appearance resembles that
of A754 where a complex merger scenario or a sloshing core seem to
be the possible explanations. Only deeper X-ray data and many
redshift measurements in a more extended cluster region could allow
us to better understand the dynamics of A1914.

\section*{Acknowledgments}

We are in debt with Federica Govoni for the VLA radio image she kindly
provided us.  M.G. acknowledges financial support from
PRININAF2010. This work has been supported by the Programa Nacional de
Astronom\'\i a y Astrof\'\i sica of the Spanish Ministry of Science
and Innovation under grants AYA2010-21322-C03-02, AYA2007-67965-C03-01
and AYA2010-21887-C04-04.

This publication is based on observations made on the island of La
Palma with the Italian Telescopio Nazionale Galileo (TNG), which is
operated by the Fundaci\'on Galileo Galilei -- INAF (Istituto
Nazionale di Astrofisica) and is located in the Spanish Observatorio
of the Roque de Los Muchachos of the Instituto de Astrofisica de
Canarias.

This research has made use of the NASA/IPAC Extragalactic Database
(NED), which is operated by the Jet Propulsion Laboratory, California
Institute of Technology, under contract with the National Aeronautics
and Space Administration.

This research has made use of archival data obtained at the
Canada-France-Hawaii Telescope (CFHT), which is operated by the
National Research Council of Canada, the Institut National des
Sciences de l'Univers of the Centre National de la Recherche
Scientifique of France, and the University of Hawaii.

This research has made use of the galaxy catalog of the Sloan Digital
Sky Survey (SDSS). Funding for the SDSS has been provided by the
Alfred P. Sloan Foundation, the Participating Institutions, the
National Aeronautics and Space Administration, the National Science
Foundation, the U.S. Department of Energy, the Japanese
Monbukagakusho, and the Max Planck Society. The SDSS Web site is
http://www.sdss.org/.

The SDSS is managed by the Astrophysical Research Consortium for the
Participating Institutions. The Participating Institutions are the
American Museum of Natural History, Astrophysical Institute Potsdam,
University of Basel, University of Cambridge, Case Western Reserve
University, University of Chicago, Drexel University, Fermilab, the
Institute for Advanced Study, the Japan Participation Group, Johns
Hopkins University, the Joint Institute for Nuclear Astrophysics, the
Kavli Institute for Particle Astrophysics and Cosmology, the Korean
Scientist Group, the Chinese Academy of Sciences (LAMOST), Los Alamos
National Laboratory, the Max-Planck-Institute for Astronomy (MPIA),
the Max-Planck-Institute for Astrophysics (MPA), New Mexico State
University, Ohio State University, University of Pittsburgh,
University of Portsmouth, Princeton University, the United States
Naval Observatory, and the University of Washington.

\end{document}